\documentclass[aps,prb,10pt,twocolumn,longbibliography,superscriptaddress]{revtex4-1}

\usepackage{graphicx}
\usepackage{amsmath}

\begin{document}

\title{Dispersive read-out of ferromagnetic resonance for strongly coupled magnons and microwave photons}

\author{J.~A.~Haigh}
\affiliation{Hitachi Cambridge Laboratory, J.~J.~Thomson Ave., Cambridge CB3 0HE, United Kingdom}
\author{N.~J.~Lambert}
\affiliation{Microelectronics Group, Cavendish Laboratory, University of Cambridge, J.~J.~Thomson Ave., Cambridge CB3 0HE, United Kingdom}
\author{A.~C.~Doherty}
\affiliation{Centre for Engineered Quantum Systems, School of Physics, The University of Sydney, Sydney, NSW 2006, Australia}
\author{A.~J.~Ferguson}
\affiliation{Microelectronics Group, Cavendish Laboratory, University of Cambridge, J.~J.~Thomson Ave., Cambridge CB3 0HE, United Kingdom}

\date{\today}

\begin{abstract}
We demonstrate the dispersive measurement of ferromagnetic resonance in a yttrium iron garnet sphere embedded within a microwave cavity. The reduction in the longitudinal magnetization at resonance is measured as a frequency shift in the cavity mode coupled to the sphere. This measurement is a result of the intrinsic non-linearity in magnetization dynamics, indicating a promising route towards experiments in magnon cavity quantum electro-dynamics. 
\end{abstract}

\pacs{}

\maketitle

\section{Introduction}

The dispersive limit of interaction between two coupled oscillators occurs when they are significantly detuned from one another, but their amplitudes and frequencies are still linked contingent on a non-linearity in some part of the system. In this coupling regime the probing of one oscillator can be used to infer the properties and state of the other \cite{haroche_exploring_2006} with a well defined and limited backaction \cite{braginsky_quantum_1995}. The use of dispersive measurement techniques has allowed photon number measurement \cite{brune_quantum_1990,brune_lamb_1994,gleyzes_quantum_2007} in cavity quantum electro-dynamics (QED), been applied to mechanical resonators \cite{Aspelmeyer_cavity_2014} coupled to both optical \cite{thompson_strong_2008} and microwave cavities \cite{regal_measuring_2008}, and has been central to the measurement of individual \cite{wallraff_approaching_2005,Schuster_ac_2005} and coupled \cite{majer_coupling_2007} superconducting qubits in circuit QED.

In this letter we demonstrate the dispersive measurement of ferromagnetic resonance (FMR) in a ferrimagnetic insulator strongly coupled to a microwave cavity \cite{roberts_magnetodynamic_1962}. This system has been the subject of recent interest in the dynamics of hybrid magnon-photon modes \cite{zhang_strongly_2014}, with the prospect of coherent control of long-lived single magnon states at low temperatures \cite{tabuchi_hybridizing_2014} and the coupling via the resonator to qubits \cite{tabuchi_coherent_2014} or other magnetic elements \cite{huebl_high_2013}.  While conventional FMR measures how the complex susceptibility tensor depends on frequency and magnetic field \cite{farle_ferromagnetic_1998}, here we measure the change in longitudinal magnetization, a quantity proportional to the magnon number. The experiment demonstrates that the methods of cavity QED can be applied in this system at room temperature, albeit in the classical limit of many magnons and photons, differentiating this work from previous measurements of longitudinal magnetization by inductive coupling \cite{bloembergen_relaxation_1954}, magneto optical methods \cite{meckenstock_kerr_2005,gerrits_direct_2007}, Brillouin light scattering \cite{demokritov_brillouin_2001} and magnetic resonance force microscopy \cite{rugar_mechanical_1992,klein_measurement_2003}. We note that dilute paramagnetic spin ensembles have previously been measured dispersively \cite{Amsuss_cavity_2011,ranjan_probing_2013} motivated by the prospect of storing quantum information in a spin ensemble \cite{Imamoglu_cavity_2009}. However, due to the absence of exchange coupling in these paramagnetic systems low temperatures and large static fields are needed to polarize the system to allow access to distinct magnon modes \cite{wesenberg_quantum_2009}.

\section{Experimental Setup}

\begin{figure}
    \centering
    \includegraphics[width=0.85\columnwidth]{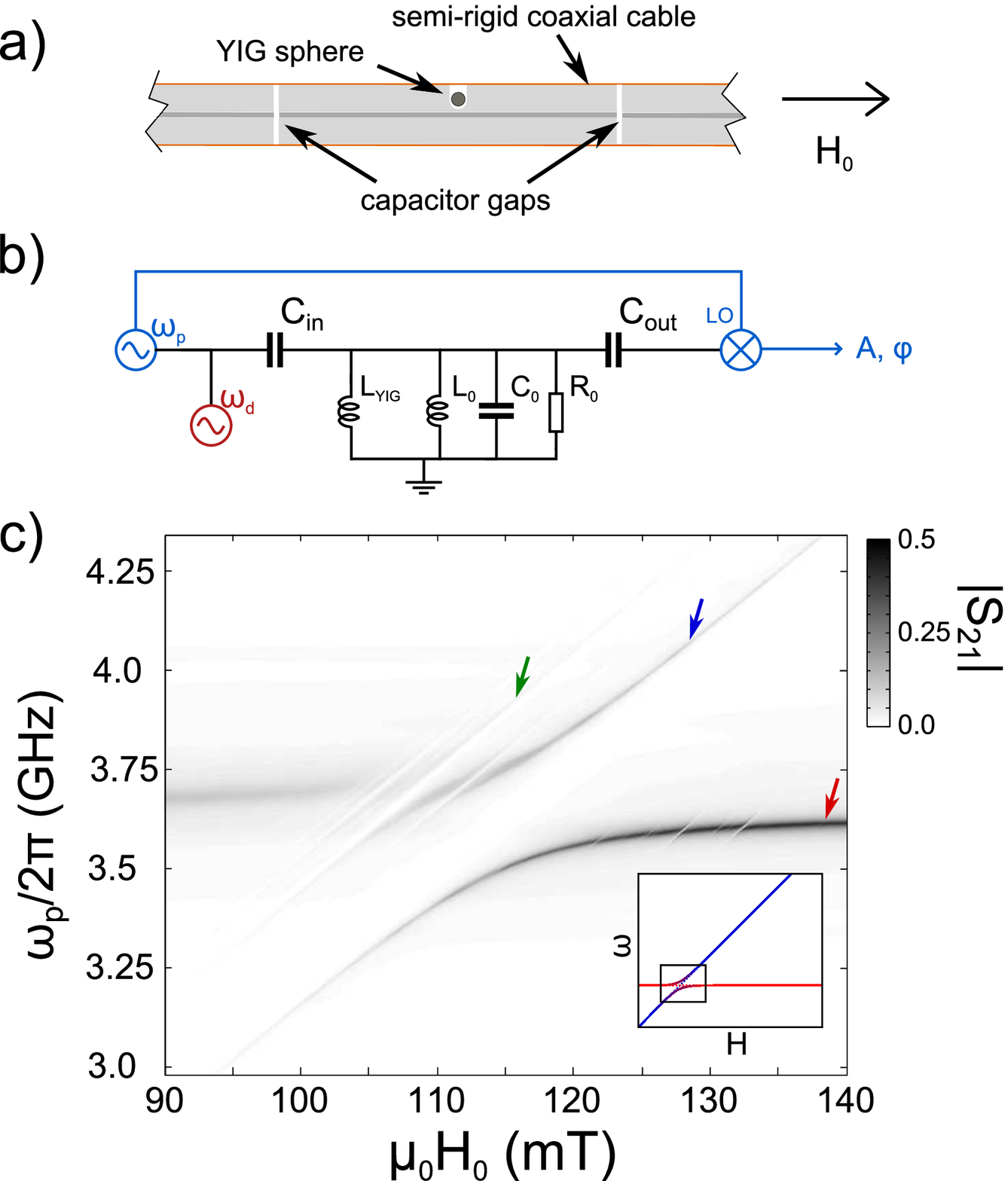}
		\caption{a) Schematic of the semi-rigid coaxial resonator with YIG sphere inside. b) Circuit equivalent diagram of the measurement setup and cavity. The probe tone $\omega_p$ and mixing are provided by a vector network analyzer (blue) while FMR drive tone $\omega_d$ is from a separate microwave source (red). c)Transmission $|S_{21}|$ through cavity as a function of magnetic field, demonstrating the strong coupling of the uniform FMR mode (blue arrow) to the $\lambda/2$ cavity mode (red arrow). In addition a number of other spin wave modes in the sphere couple to the cavity; the strongest is indicated by the green arrow. Inset shows a reference map of the measurement region over the modes of the system (only the uniform FMR mode is shown).}
		\label{s_c}
\end{figure}

Our microwave cavity is constructed from a 28\,mm length of semi-rigid copper coaxial cable of diameter 3.5\,mm (fig. \ref{s_c}(a)), chosen to give a resonance at around 3.5\,GHz for the fundamental mode. Capacitive couplings to the input and output of the resonator are achieved via a $<1$\,mm gap between the inner conductor of the cavity and two cables which connect the device to the measurement instruments. To ensure that the cavity is symmetric and undercoupled the capacitor gaps are adjusted at each end while measuring the reflection coefficients $S_{22}$ and $S_{11}$, and we achieve a loaded $Q$-factor of $260$ which can be compared to an internal $Q$-factor of $515$. The 1\,mm diameter YIG sphere is embedded within the dielectric of the semi-rigid cable, at the antinode of magnetic field.
The cavity is mounted in an electromagnet, with its long axis parallel to the applied field direction. Transmission measurements are made with a vector network analyzer, with an additional microwave source coupled in via a power combiner to allow two-frequency measurements (fig.\,\ref{s_c}(b)).

Strong coupling of the uniform ferromagnetic resonance mode to the fundamental mode of the electromagnetic resonator is demonstrated by a measurement of the transmission amplitude $\left|S_{21}\right|$ as a function of magnetic field and probe frequency $\omega_p$ (fig.\,\ref{s_c}(c)). The characteristic frequency splitting of the coupled modes observed at the degeneracy point gives a coupling rate of $g/2\pi=130$\,MHz, greater than the linewidth of both the cavity mode and the FMR mode ($\kappa/2\pi=14$\,MHz and $\gamma/2\pi=6$\,MHz respectively). This indicates that the strong coupling regime $g\gg\kappa,\gamma$ is achieved, with a co-operativity ($g^2/\kappa\gamma$) of around 200. In addition to the uniform mode, other non-uniform magneto-static modes in the sphere \cite{walker_resonant_1958,fletcher_ferrimagnetic_1959} couple to the resonator due to the radial dependence of microwave field in the cavity. One of these, indicated by the green arrow in the figure, with a coupling rate of $\approx35$\,MHz, is also strongly coupled to the resonator.

\section{Dispersive measurements}

\begin{figure}
    \centering
    \includegraphics[width=0.85\columnwidth]{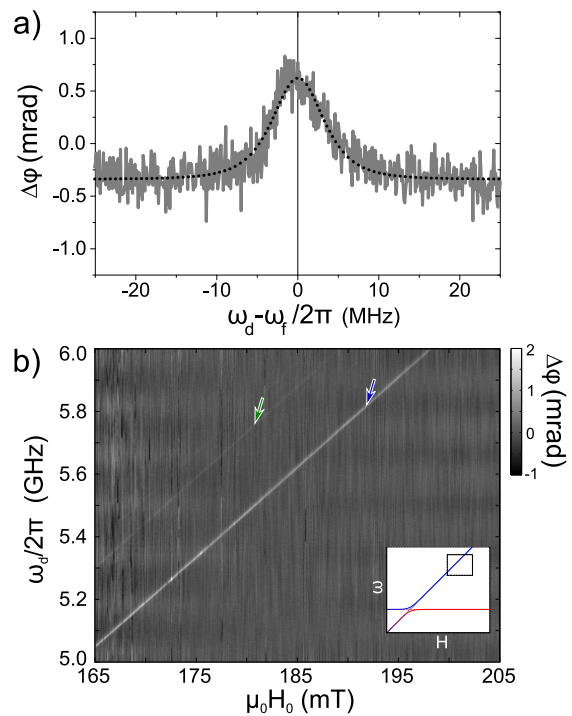}
		\caption{Dispersive measurement: a) Phase shift of the probe tone $\omega_p$ (-10\,dBm) at the resonant frequency of the cavity as a function of the drive frequency $\omega_d$, for input drive power 25\,dBm (off resonance with the cavity a large amount of power is reflected). The field is fixed at $\mu H_0=1800$\,G. A slow time dependent dependent phase drift and field independent background have been subtracted. The dashed line is a fit to the square of a Lorentzian (square of eqn.\,\ref{eq:b}), appropriate since we are sensitive to the square of the transverse magnetization. The fit gives the magnon decay rate $\gamma/2\pi=6$\,MHz. b) Phase shift as a function of magnetic field and drive frequency. Probe frequency tracks the resonance frequency of the cavity, $\approx 3.55$\,GHz, as a function of field.}
		\label{dispers}
\end{figure}

To perform the dispersive measurement we move away from the degeneracy point by tuning the static magnetic field so that the FMR mode and cavity mode are not resonantly coupled. Two microwave tones are applied to the cavity. We monitor the phase of a transmitted probe tone $\omega_p$ close to the cavity resonance $\omega_c$ as a drive tone $\omega_d$ is swept across the resonant frequency of the FMR mode. As shown in fig.\,\ref{dispers}(a), when the drive tone is resonant with the FMR mode there is an increase in the phase of probe tone of nearly $1$\,mrad, corresponding to a cavity frequency shift of $+5$\,kHz. This measurement is repeated for different external magnetic fields (fig. \,\ref{dispers}(b)). The two FMR lines which couple most strongly to the cavity are visible and the field dependence of the uniform mode \cite{kittel_theory_1948} yields the gyromagnetic ratio $\gamma=28.6$\,GHz~T$^{-1}$.

\begin{figure}
    \centering
    \includegraphics[width=0.8\columnwidth]{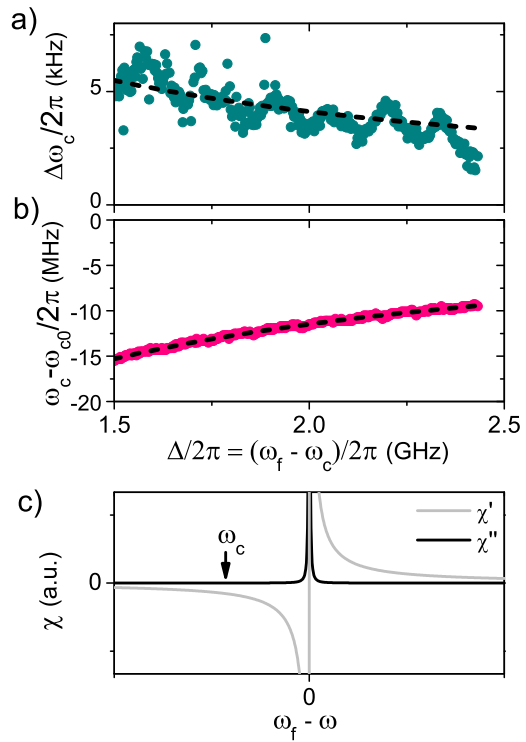}
		\caption{Dependence of measurements on detuning between cavity and FMR mode frequencies. a) change in cavity frequency due to FMR, taken from peak amplitude in phase change plotted in fig. \ref{dispers}(a). b) change in cavity frequency due to the equilibrium susceptibility of the sphere from that when $\chi=0$. The value of $\omega_c0$ is obtained from the offset in the $1/\Delta$ fit. c) For reference, the calculated general form of dispersive and dissipative susceptibility of the sphere as a function of detuning from the FMR resonance, from eqn.\,\ref{eq:a},\ref{eq:b}, for low damping $\alpha$.}
		\label{delta_omegas}
\end{figure}

The variation with static magnetic field is used to study the dispersive measurement as a function of detuning between cavity and FMR mode frequencies, $\Delta=\omega_f-\omega_c$. The change in the cavity frequency when the drive tone is applied, $\Delta\omega_c$, is computed from the measured phase change in $\omega_p$ and follows a $1/\Delta$ dependence as indicated by the dashed fitted line in fig.\,\ref{delta_omegas}(a). We also measure the equilibrium cavity frequency shift with the drive tone turned off, obtained from the peak in transmission amplitude as a function of the probe frequency, also well fitted by $1/\Delta$ (fig.\,\ref{delta_omegas}(b)).

The dispersive measurement can be understood by considering the susceptibility $\chi=\chi'+i\chi''$ transverse to a large static field of a uniformly magnetized ferromagnetic sphere. In the small amplitude limit \cite{chikazumi_physics_1997},
\begin{equation}
\chi' = \frac{\gamma M_0 (\omega_f - \omega)}{(\omega_f - \omega)^2+\omega_f^2\alpha^2}
\label{eq:a}
\end{equation}
\begin{equation}
\chi'' = \frac{\alpha \gamma M_0 \omega_f }{(\omega_f - \omega)^2+\omega_f^2\alpha^2}
\label{eq:b}
\end{equation}
where $\omega_f=\gamma H_0$ is the resonant frequency, $\gamma$ is the gyromagnetic ratio and $\alpha$ is the Gilbert damping coefficient. For large detuning $\Delta=\omega_f - \omega_c\gg\omega_f\alpha$ the susceptibility is dominated by the dispersive (real) part, and reduces to
\begin{equation}
\chi \sim \frac{\gamma M_0 }{\Delta}.
\label{eq:c}
\end{equation}

The change in susceptibility during magnetization precession is taken in to account through the change in the projection of the magnetization along the static field direction. The condition that $|M_z|^2+|M_y|^2+|M_x|^2=M_0^2$ gives $|M_z|\approx M_0 \left(1 - \frac{|M_x|^2+|M_y|^2}{2 M_0^2}\right)$ for	$M_z\gg M_x,M_y$ \cite{gurevich_chapter_1966}. Replacing $M_0$ with $M_z$ in eqn.\,\ref{eq:c} results in a reduction in the dispersive susceptibility by a factor $M_z/M_0$.
Cavity perturbation theory \cite{pozar_microwave_2004} indicates that a small change in some part of the material properties of the cavity results in a proportional change in frequency, in our case the only modification is that of the permeability $\mu=\mu_{0}(1+\chi)$ of the sphere \cite{gurevich_ferrites_1963}. The equilibrium cavity shift from its bare value $\omega_{c0}$ for $\chi=0$ is then proportional to the susceptibility given by eqn. \ref{eq:c}. When we additionally drive FMR, the reduction of this susceptibility by $M_z/M_0$ gives rise to the frequency shift $\Delta\omega_c$, which then also follows a $1/\Delta$ dependence as observed in the experiment.

\begin{figure}
    \centering
    \includegraphics[width=0.8\columnwidth]{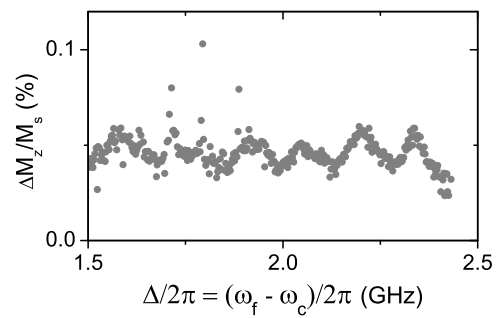}
		\caption{Percentage change in longitudinal magnetization $M_z$ as a function of detuning.}
		\label{mzgraph}
\end{figure}

From our measurement we obtain an estimate of the change in $M_z$ through comparison of the two frequency shifts discussed:
\begin{equation}
\frac{\Delta\omega_c}{\omega_c-\omega_{c0}} \approx \frac{\Delta\chi}{\chi} \approx \frac{\Delta M_z}{M_0}
\end{equation}
This results in the data plotted in fig.\,\ref{mzgraph}, where it is seen that the amplitude of the change in $M_z$ is independent of detuning.

\begin{figure}
    \centering
    \includegraphics[width=0.8\columnwidth]{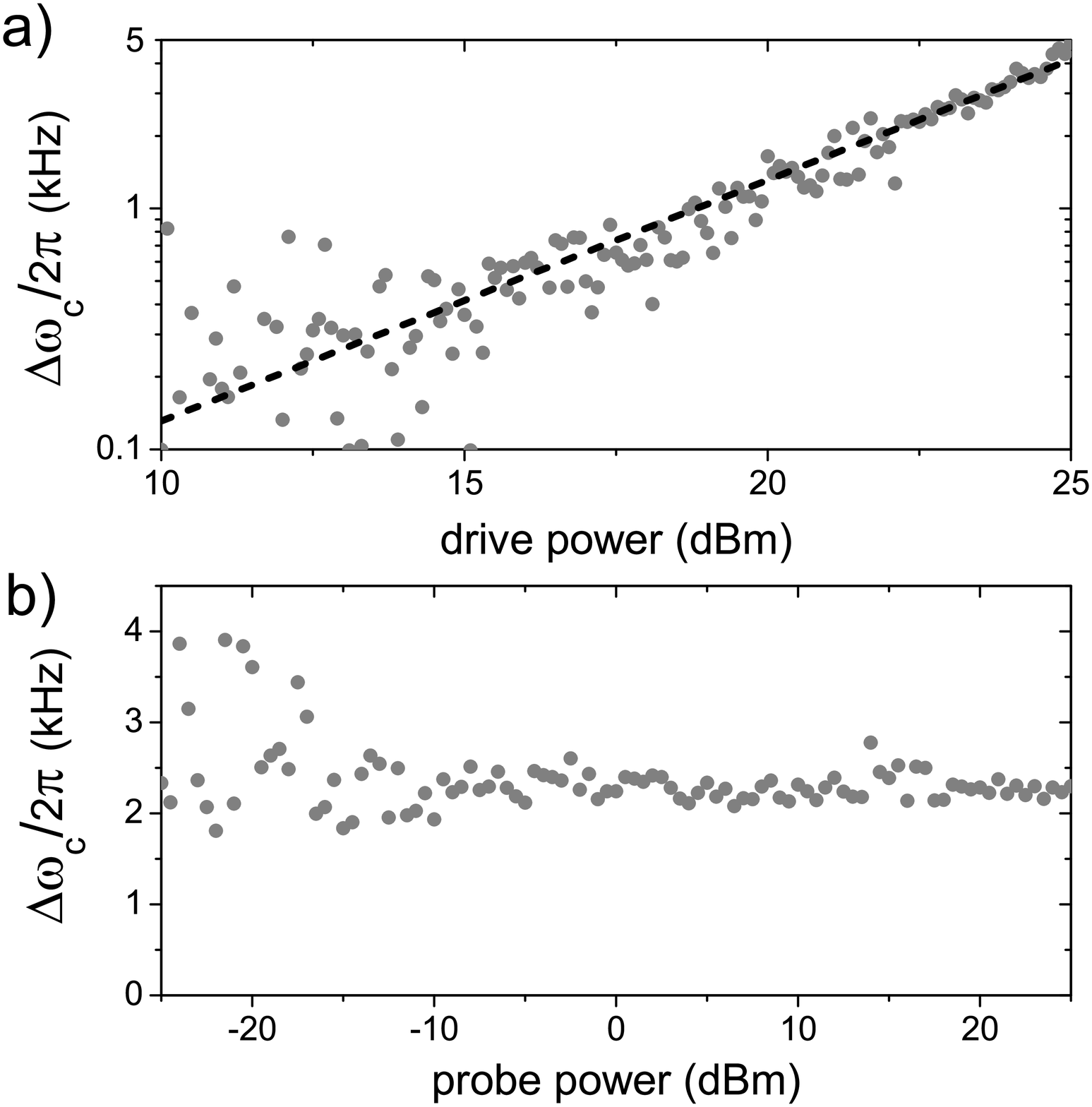}
		\caption{a) Dependence of the cavity frequency shift on the FMR driving microwave power at fixed static magnetic field $\mu_0H_0=180$\,mT. A linear fit to the data is shown. b) Power dependence of the cavity frequency shift as a function of probe power at the cavity resonance frequency for the same static field.}
		\label{power}
\end{figure}

Figure \ref{power}(a) shows the peak frequency shift as a function of the power of the FMR drive tone. The shift is linearly proportional to the power, and therefore quadratic in the magnetic field amplitude. This is due to the fact that we are sensitive to the square of the transverse magnetizations, which are themselves linear in the applied r.f. field as given by the susceptibility relation. Figure\,\ref{power}(b) shows the dispersive measurement as a function of the power of the probe tone. Here there is no dependence of the amplitude of the frequency shift over a wide range of powers, with the only change being in the signal to noise ratio improving as the probe field is increased.

\section{Discussion}

Recent work on magnon-photon interactions \cite{huebl_high_2013,zhang_strongly_2014} have analyzed this system in terms of a fully quantum mechanical Hamiltonian \cite{soykal_strong_2010}. As such, it is useful to discuss a description of our measurements based on this model. In the dispersive limit $\Delta\gg g$, degenerate second order perturbation theory gives an effective Hamiltonian analogous to circuit QED \cite{blais_cavity_2004}.  In the rotating frame at the drive frequency $\omega_d$ we obtain
\begin{multline}
H_{\mathrm{eff}} = \Delta_f J_z + \hbar\Delta_c a^{\dagger} a + g_0 (\beta J^+ +\beta^* J^-) \\ + \hbar\frac{2 g_0^2}{\Delta} a^{\dagger} a J_z +  \frac{g_0^2}{\Delta}(J_z + \frac{1}{\hbar}(J_x^2+J_y^2)) 
\label{eq:H}
\end{multline}
where $\Delta_f = \omega_f-\omega_d$ is the drive detuning from the FMR mode, $\Delta_c=\omega_c-\omega_d$ is the drive detuning from the cavity and $J_z$ is the large angular momentum associated with the magnetization $M_z$. $J^+$ and $J^-$ are the corresponding angular momentum raising and lowering operators and $\beta$ is proportional to the amplitude of the microwave drive at frequency $\omega_d$. Here we use the single spin coupling rate $g_0$ related to the measured total coupling by $g=g_0\sqrt{N}$. The first two terms are the bare uniform ferromagnetic mode and cavity mode with lowering operator $a$.  The third represents the microwave driving of the FMR mode with amplitude $\beta$ and frequency $\omega_d$. The last two terms are the ones that we are interested in: a modification of the cavity frequency proportional to $J_z$ and a magnon non-linearity respectively. It is clear that, as the cavity frequency shift term commutes with the FMR mode Hamiltonian, this measurement of $J_z$ is in principle quantum non-demolition \cite{caves_measurement_1980} and as in the classical analysis depends inversely on the detuning.
 
In our measurements, the change in the number of magnons \cite{turov_chapter_1966} $n$ from the equilibrium thermal occupation ($M_z = M_0-\mu_B n$) when the system is driven is $n\approx3\times10^{15}$, where we have used an effective value for the number of spins at room temperature to calculate $M_0$ \cite{Gilleo_magnetic_1958}. While this magnon number is large, micro-fabricated on-chip devices, enabling substantial reduction in volume while maintaining strong coupling \cite{huebl_high_2013} could, at low temperatures, make dispersive measurements in the low $n$ limit and the prospect of the generation of magnon number states through measurement \cite{brune_manipulation_1992} a possibility. In addition, the ability to study non-linear effects in this system without the need to introduce a qubit, as has been recently achieved by Tabuchi et al. \cite{tabuchi_coherent_2014}, instead exploiting the intrinsic anharmonicity in the magnetic system itself, promises to enable further experiments in magnon cavity QED.

\section{acknowledgements}
We would like to acknowledge support from Hitachi Cambridge Laboratory, and EPSRC Grant No. EP/K027018/1. AJF is supported by a Hitachi Research fellowship. ACD is supported by the ARC via the Centre of Excellence in Engineered Quantum Systems (EQuS), project number CE110001013. JAH is grateful to Andrew Ramsay for useful discussions.

\bibliography{bibliography}

\bibliographystyle{apsrev}

\end{document}